\documentclass[fleqn,10pt,letterpaper,twocolumn]{article} 
\usepackage{ol2}
\usepackage[draft]{hyperref}
\usepackage[fleqn]{amsmath}
\usepackage{color}

 \newcommand {\e}{{\rm e}}
\renewcommand {\Re}{\mathop{\mathrm{Re}}\nolimits}
\newcommand {\rmi}{{\rm i}}
\begin{document}

\twocolumn[ 

\title{Spontaneous emission enhancement in metal-dielectric metamaterials}

\author{Ivan Iorsh$^{1,2,*}$, Alexander Poddubny$^{1,3}$, Alexey Orlov$^{1}$,  Pavel Belov,$^{1,4}$, and Yuri S. Kivshar$^{1,5}$}

\address{
$^1$St. Petersburg University of Information Technologies, Mechanics and Optics (ITMO), St. Petersburg 197101, Russia\\
$^2$St. Petersburg Academic University — Nanotechnology Research and Education Centre, St. Petersburg 194021, Russia\\
$^3$Ioffe Physical Technical Institute, St. Petersburg 194021, Russia\\
$^4$ School of Electronic Engineering and Computer Science, Queen Mary University of London, London E1 4NS, U.K. \\
$^5$ Nonlinear Physics Center and Centre for Ultrahigh-bandwidth Devices for Optical Systems (CUDOS), \\Research School of Physics and Engineering,
Australian National University, Canberra ACT 0200, Australia\\
$^*$Corresponding author: iorsh86@yandex.ru
}

\begin{abstract}
We study the spontaneous emission of a dipole emitter imbedded into a layered metal-dielectric metamaterial. We demonstrate
ultra-high values of the Purcell factor in such structures due to a high density of states with hyperbolic isofrequency
surfaces. We reveal that the traditional effective-medium approach greatly underestimates the value of the Purcell factor
due to the presence of an effective nonlocality, and we present an analytical model which agrees well with numerical calculations.
  \end{abstract}

\ocis{160.1190, 160.3918.}

 ] 

\noindent

Spontaneous emission rate of a light source can be tuned by engineering its environment. A ratio of the decay rate in the media as compared to that in vacuum 
is usually termed as the Purcell factor~\cite{Purcell}, and the possibility of a controllable change of of the radiative lifetime has been demonstrated in 
various optical systems, including photonic crystals and plasmonic nanostructures~\cite{Yablonovich1987,Hughes2007PRB,Jun2008prb}.

In metamaterials the Purcell factor can be greatly enhanced, with two possible mechanisms of achieving high values of the Purcell factors. In the first case, 
we should place a dipole emitter into a metamaterial that is described as an uniform hyperbolic medium. Such hyperbolic media are uniaxial media characterized by
the permittivity tensor of the diagonal form with the principal components being of the opposite signs~\cite{Xie2009,narimanov2010}. The density of photonic states 
in such a system diverges. As a result, a huge Purcell factor can be reached, and its values being determined either by losses and inhomogeneity of 
the medium~\cite{narimanov2010} or by a spatial extent of the source~\cite{PoddubnyHyperb2011}. The second mechanism of achieving high Purcell factors 
employs the excitation of surface plasmon polaritons (SPPs) at the metal-dielectric interfaces inside a metamaterial~\cite{Jun2008prb,Sipe2011}.

\begin{figure}[htb]
\centerline{\includegraphics[width=6.5cm]{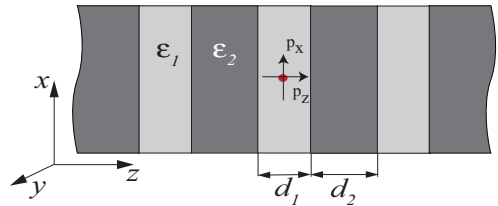}}
\caption{Geometry of the layered metamaterial formed by alternating
metal and dielectric layers. Arrows show the orientations of a dipole emitter (a red filled circle).}\label{fig1}
\end{figure}

In this Letter, we study the spontaneous emission of a dipole emitter imbedded into a layered metal-dielectric metamaterial (see Fig.~\ref{fig1}) and reveal
the possibility for ultrahigh values of the Purcell factor due to a high density of states with hyperbolic isofrequency
surfaces. In particular, we demonstrate the dramatic dependence of the Purcell factor near the SPP resonance on the ratio of the layer thicknesses and the 
dipole orientation, so that the Purcell factor can have either maximum or minimum at the plasmon frequency for high and low metal filling factors, respectively.
This revealed mechanism of the enhancement of the Purcell factor has an interface nature, and thus it can not be described within any 
traditional homogenization approach. However, here we present an analytical model which agrees well with our numerical results.

The hyperbolic medium at certain frequency ranges can be realized by layered metal-dielectric nanostructures. Through conventional 
effective-medium approach a metal-dielectric nanostructure formed by pairs of layers with permittivities $\varepsilon_1, \varepsilon_2$, 
and thicknesses $d_1, d_2$ (see Fig.~\ref{fig1}) can be modeled as an uniaxial anisotropic medium with permittivity tensor of the form:
\[
\varepsilon_\mathrm{eff} = \left( \begin{matrix}
\varepsilon_\bot & 0 & 0\\0 & \varepsilon_\| & 0\\0 & 0 & \varepsilon_\|
\end{matrix}\right), \
\begin{array}{lcl}
\varepsilon_\| = \dfrac{\varepsilon_1 d_1 + \varepsilon_2 d_2}{d_1 + d_2},
\\
\varepsilon_\bot = \left(
\dfrac{\varepsilon_1^{-1} d_1 + \varepsilon_2^{-1} d_2}{d_1 + d_2}
\right)^{-1}
\end{array}
\]
At the same time, one of the key properties of such a medium is its strong nonlocality due to excitation of SPP modes at metal-dielectric interfaces~\cite{orlov2011arXiv}.

Such structures were a subject of extensive theoretical and experimental studies being suggested for many applications such as superlenses with subwavelength resolution~\cite{Belov2006}, a simple realization of the so-called hyperlens~\cite{Hyper2007} and optical nanocircuits~\cite{metactronics}, as well as 
invisibility cloaks~\cite{Cai2008}. For a certain ratio of the layer thicknesses, one can reach the hyperbolic regime when high Purcell factors can be expected~\cite{Sipe2011}.
In particular, it was demonstrated that the effective medium approach is not always applicable for the calculation of the radiation emission, and it can overestimate
the Purcell factor~\cite{Sipe2011}. In the regime we consider here the effective-medium approach underestimates the Purcell factor, so that the high 
Purcell factors can be attained even in the structures which are not hyperbolic when being treated as effective media.
 
We study a system shown schematically in Fig.~\ref{fig1}. A point dipole emitter is placed in the center of a dielectric layer of the periodic metal-dielectric 
layered nanostructure, the simplest version of a metamaterial. In such a planar system, we consider independently two orientations of the dipole: 
parallel ($p_{x}$) and perpendicular ($p_{z}$) to the interface planes. Although the system is periodic, it can be formally treated as a cavity, where the 
central dielectric layer (labeled as the layer 1) with the dipole is surrounded by two identical mirrors. One can employ then the standard results 
of the cavity theory~\cite{Martini1990} to calculate the Purcell factors $F_p^{x}=1+R_{x}$, and $F_p^{z}=1+R_{z}$, where
\begin{subequations}\label{Purcell}
\begin{align}
&R_{x}=\frac{3}{2}\Re \int\limits_0^{\infty}\frac{dk_{\parallel}k_{\parallel} }{k k_{z}^{(1)}} \left[ \frac{r_{s}}{(1-r_{s})} +  \frac{r_{p} (k_z^{(1) })^2 }{k^2(1-r_{p})} \right] \\
&R_{z}=-3\Re \int\limits_0^{\infty}\frac{dk_{\parallel}k_{\parallel}^3 }{k_{z}^{(1)} k^3} \frac{r_{p}}{(1+r_{p})} \:,
\end{align}\label{Purc}
\end{subequations}
where $k=\sqrt{\varepsilon_{1}}\omega/c$ and  $k_{z}^{(1,2)}=\{\varepsilon_{1,2}\omega^{2}/c^{2}-k_{\parallel}^2\}^{1/2}$
Integration in Eqs.~\eqref{Purc} is performed over the in-plane component of the wavevector $k_{\parallel}$.
 The amplitude reflection coefficients of the semi-infinite periodic mirrors for {$s$} and {$p$} polarizations,
$r_{s}$  and $r_p$ are readily calculated
(see, e.g.~\cite{yariv2002})
\begin{align}
r^{s,p}=\frac{r_{1}^{s,p}} {1-t_{1}^{s,p} \e^{\rmi k_{\perp}D} }\e^{\rmi k_{z}^{(1)}d_1}
\end{align}
where $r_1$, $t_1$ are the reflection and transmission coefficients for one period and a given polarization, 
$D=d_1+d_2$ is the structure period, and $k_{\perp}$ is the Bloch wavevector defined by the dispersion relation:
\begin{align}
\label{disp}
\cos(k_{\perp}D)=&\cos(k_{z}^{(1)}d_1)\cos(k_{z}^{(2)}d_2)- \\
&\frac{1}{2}\left(\frac{Z_1^{s,p}}{Z_2^{s,p}}+\frac{Z_2^{s,p}}{Z_1^{s,p}}\right)\sin(k_{z}^{(1)}d_1)\sin(k_{z}^{(2)}d_2)\:,\nonumber
\end{align}
where $Z_i^s=k_{z}^{(i)},\quad Z_i^p={\varepsilon_i}/{k_{z}^{(i)}}, i=1,2$. In our numerical calculation we chose silver as metal with
the dielectric constant described by the Drude model~\cite{JohnsonChristy}: $\varepsilon_{2}=\varepsilon_{\infty}-{\omega_p^2}/{\omega(\omega+\rmi\gamma)}$
with $\varepsilon_{\infty}\approx 4.96,\:\omega_p\approx 8.98~{\rm eV}, \gamma\approx 0.018~{\rm eV}$, while the dielectric constant $\varepsilon_{\rm 1}$ was set to unity.  
The structure period is fixed to $D=30$~nm, and only a ratio of the thickness of silver and dielectric layers is varied, $\eta=d_{\rm 2}/d_{\rm 1}$.

\begin{figure}[htb]
\centerline{\includegraphics[width=7.5cm]{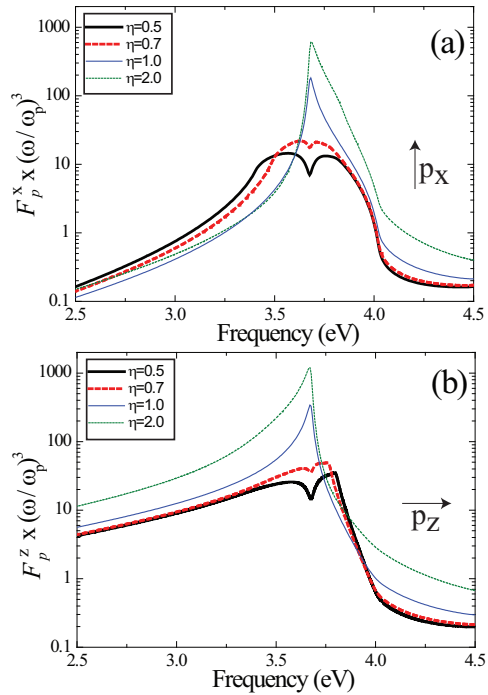}}
\caption{Purcell factor for different layer thicknesses vs. frequency. (a,b) Two orientations of the dipole emitter:
parallel and perpendicular to the layers, respectively.}
\label{fig2}
\end{figure}

The dependence of Purcell factors~\eqref{Purc} on the frequency for different metal thicknesses is shown in Fig.~\ref{fig2}.
For better presentation, the Purcell factor is scaled by cube of the light frequency, since the vacuum decay rate $\tau_{\rm vac}^{-1}$ is proportional to $\omega^3$~\cite{Kavokin}.
We observe that for the case when the silver layer is thicker than the dielectric layer, there is {\em a sharp maximum} of the Purcell factor at the frequency of 
the SPP resonance where the condition $\Re\varepsilon_{\rm 2}=-1$ is fulfilled. However, for the case of {\em a wider dielectric layer}, the picture changes dramatically,
 and we observe {\em a local minimum} at the frequency of the SPP resonance. Next, we examine the latter case in detail.
 
\begin{figure}[t]
\centerline{\includegraphics[width=7.3cm]{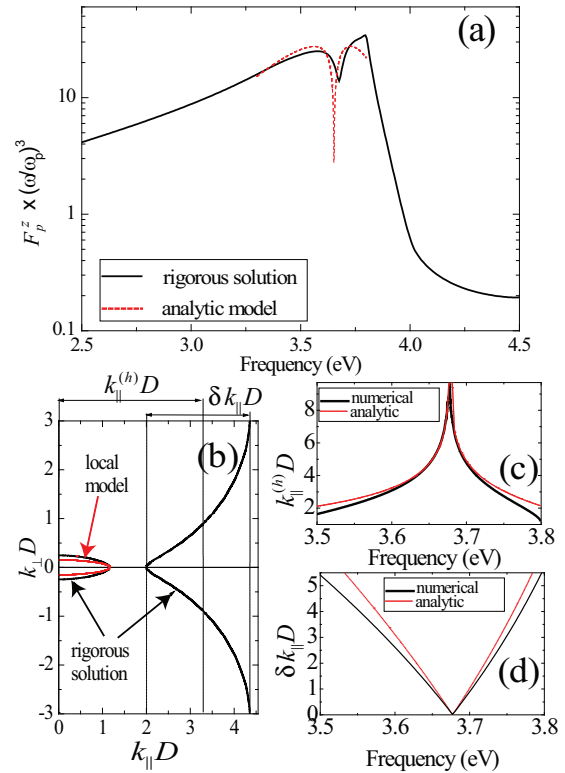}}
\caption{(a) Purcell factor of a dipole oriented perpendicular to the layers for $\eta=0.5$. Red line corresponds to expression~\eqref{analyt} ; (b) Isofrequency contours for the eigenmodes of the structure, $\hbar\omega=3.57$ eV.
 (c) Position of the hyperbolic contour vs. frequency. Red line corresponds to expression~\eqref{analyt2} ;  (d) Width of the hyperbolic contour vs. frequency. Red line corresponds to expression~\eqref{analyt2}.} 
 \label{fig3}
\end{figure}

The solid line in Fig.~\ref{fig3}(a) depicts the Purcell factor calculated for the in-plane 
dipole for $\eta=0.5$. These results reveal two characteristic effects. First,  the Purcell factor can be enhanced even 
in the non-hyperbolic case. Second, the Purcell factor exhibits a sharp dip near the SPP resonance frequency.
To explain this puzzling behavior,  in Fig.~\ref{fig3}(b) we plot the isofrequency contours $k_{\perp}(k_{\parallel})$ for the close-to-resonance frequencies.
We observe two distinct contours: an elliptic contour $k_{\perp}^2/\varepsilon_{\perp}+k_{\parallel}^2/\varepsilon_{\parallel}=(\omega/c)^2$ predicted by the effective-medium theory,  and an additional hyperbolic contour. The importance of this contour has been mentioned in Ref.~\cite{orlov2011arXiv}, where the isofrequency contours of metal-dielectric nanostructures were studied by the exact and effective-medium approach, and they attributed to the effect of nonlocal response of such layered structures.  
It is this extra contour that determines a dramatic enhancement of the Purcell factor in Fig.~\ref{fig2} 
in the regime when the structure is not hyperbolic.

Next, we derive a closed form expression for the corresponding contribution to the Purcell factor,
\begin{equation}
\label{analyt}
F_p^z =\frac{3\pi}{8}\frac{(k_{\parallel}^{(h)})^2\delta k_{\parallel}}{(\omega/c)^3},
\end{equation}
\begin{equation}
\label{analyt2}
 k_{\parallel}^{(h)}=\frac{1}{d_1}\ln\xi, \;
 \delta k_{\parallel}=\xi^{-\frac{d_{\rm 1}-d_{\rm 2}}{2d_{\rm 1}}}, \;
\xi=-\frac{\varepsilon_{1}\varepsilon_{2}}
{(\varepsilon_{1}+\varepsilon_{2})^2}, 
\end{equation}
which is valid in the vicinity of the plasmonic resonance, so that $\xi\gg 1$.
Here $k_{\parallel}^{(h)}$ is the center and $\delta k_{\parallel}$ is the width of the hyperbolic contour. 
Naturally, the Purcell factor is determined by the density of states $(k_{\parallel}^{(h)})^2\delta k_{\parallel}$, and the extra factors 
in Eq.~\eqref{analyt} depend on the the orientation and  position of a dipole in the dielectric layer. The approximate expression \eqref{analyt}, 
shown in Fig.~\ref{fig3}(a) reproduces well the Purcell factor calculated numerically.

The position of the center of the hyperbolic contour, found numerically from the equation $\cos[k_{\perp}(k_{\parallel}^{(h)})D]=0$, is plotted in Fig.~\ref{fig3}(c), and
it is well described by the analytical expression \eqref{analyt}. The width of this contour as a function of the frequency is also reproduced by the function $\delta k_{\parallel}$ in Eq.~\eqref{analyt} [see Fig.~\ref{fig3}(d)]. Thus, this results uncover the origin of the dip in the Purcell factor at the surface plasmon frequency: while the position of the hyperbolic contour increases logarithmically when $\xi\to \infty$ as we approach the plasmon frequency, the width of the hyperbolic contour demonstrates a power-low decrease. Therefore, as a product of these two values the Purcell factor should vanish at the SPP resonance, but it remains finite due to losses in silver. It is also worth mentioning that the effect of the enhancement of the Purcell factor demonstrated here is purely nonlocal, since the effective-medium approach can predict only the first elliptic contour 
[dashed line in Fig.~\ref{fig3}(b)] and thus the resulting integral \eqref{Purcell} will be small.

In conclusion, we have studied the spontaneous emission process and radiation rate of a dipole imbedded into a layered metal-dielectric 
metamaterial. We have demonstrated that the effective-medium approximation underestimates the radiative decay rate in such structures
which is greatly increased in comparison with a bulk dielectric or metal. We have predicted, both analytically and numerically,
ultrahigh values of the Purcell factor in such metal-dielectric structures due to a high density of states with hyperbolic isofrequency
surfaces.

This work was supported by the Ministry of Education and Science of the Russian Federation, RFBR
and Dynasty Foundation (Russia), EPSRC (UK), and the Australian Research Council.
The authors acknowledge useful discussions with C.R. Simovski and J. Sipe.

\end{document}